\documentclass[showpacs,preprintnumbers,amsmath,amssymb,twocolumn]{revtex4-1}
\usepackage{graphicx}
\usepackage{dcolumn}
\usepackage{bm}
\usepackage{subfigure}
\usepackage{float}
\usepackage{multirow}
\usepackage{booktabs}
\usepackage{amsmath}
\usepackage{latexsym,amssymb}
\usepackage[mathscr]{eucal}
\usepackage[switch*]{lineno}
\usepackage[bookmarks=true,
   colorlinks=true,
   linkcolor=blue,
   urlcolor=blue,
   citecolor=blue,
   bookmarks=true,
   hyperindex=true
]{hyperref}

\begin{document}

\title{The gravitational baryogenesis and a new higher-order extended uncertainty principle with parameter adaptability for the minimum length}

\author{Song-Shan Luo\textsuperscript{1,2}}
\author{Zhong-Wen Feng\textsuperscript{1,2}}
\altaffiliation{Email: zwfengphy@163.com}

\vskip 0.5cm
\affiliation{1 School of Physics and Astronomy, China West Normal University, Nanchong, 637009, China
\\ 2 Institute of Theoretical Physics, China West Normal University, Nanchong, 637009, China}

\begin{abstract}
In this manuscript, we explore the baryon asymmetry of the universe by employing a novel higher-order extended uncertainty principle (EUP)  that maintains a minimum length ${\rm{\Delta }}{x_{\rm min}} =4\sqrt {\left| {\rm{\beta_0 }} \right|}\ell _p $  for both positive and negative deformation parameters. Our results demonstrate that the influence of the EUP noticeably modifies the Friedmann equations, leading to a transformation in the characteristics of the pressure and density of the Universe, and subsequently disrupting its thermal equilibrium. Additionally, by amalgamating the adapted Friedmann equations with the conventional theory of gravitational baryogenesis, one can derive a non-zero factor of baryon asymmetry $\eta$, indicating that the quantity of matter in the universe surpasses that of antimatter. Finally, we also utilized astronomical observations to constrain the bounds for both the positive and negative deformation parameters.
\end{abstract}
\maketitle
\section{Introduction}
\label{Int}
Advancements in the field of physics have revealed that nature encompasses not just regular matter, but also its counterpart, antimatter. In the standard model of particle physics, it was previously hypothesized that an equal quantity of matter and antimatter existed. However, numerous astronomical observations, such as the Big Bang Nucleosynthesis (BBN) \cite{1}, high-precision measurements of the Cosmic Microwave Background (CMB) \cite{2}, and the Wilkinson Microwave Anisotropy Probe (WMAP) experiment, have revealed the opposite, our Universe is predominantly composed of matter with only a minuscule amount of antimatter. This has resulted in an enigmatic puzzle in particle cosmology, known as the baryon asymmetry of the universe (BAU) \cite{3}.

Although the exact mechanism behind BAU remains elusive, researchers have endeavored to formulate numerous theories in an attempt to elucidate this phenomenon, such as electroweak baryogenesis \cite{4}, gravitational waves baryogenesis \cite{5}, and so on. Of all these theories, the gravitational baryogenesis introduced by Davoudiasl \cite{6} is noteworthy, as it proposes that a gravitational interaction can dynamically break CPT in an expanding Universe and lead to baryon asymmetry. The theory of gravitational baryogenesis reveals gravitational coupling as a means of generating baryon asymmetry and has attracted widespread attention and research. For instance, in the $f(R)$-theory of gravity  \cite{7}, the baryon asymmetry in the universe is explained by the gravitational baryogenesis. In Ref.~\cite{8}, the gravitational baryogenesis is studied in the framework of $f(P)$ gravity, and the applicability of such modified gravity in solving the cosmic baryon asymmetry is obtained.

Despite the significant role played by gravitational baryogenesis in comprehending BAU, it still has some shortcomings, the most notable of which is its inability to fulfill all of the Sakharov conditions. It is widely acknowledged that the universe displays baryon asymmetry only if it fulfills the Sakharov conditions \cite{9}, namely, (1) interactions that violate baryon-number conservation; (2) CP and C violation; and (3) a deviation from thermal equilibrium. In the original theory of gravitational baryogenesis, the first two Sakharov conditions are satisfied by introducing a coupling term that couples space-time to the baryon current. However, the theory fails to fulfill the third condition as it assumes thermal equilibrium, which violates the third Sakharov condition. Additionally, the original theory is incapable of generating a baryon asymmetry in the radiation-dominated universe. In order to fix this issue, some approaches have been proposed, see e.g., Refs.~\cite{10,11,12,13,14,15,16,17,18}. In particular, considering that the effect of generalized uncertainty principle (GUP hereafter, which is a kind of quantum gravity (QG) model~\cite{19,20,21,21++,a1,a2,a3,a4,b1,b2,g2,g3,g4}) can modify the properties of the universe and  change its thermodynamic state, Das, Vagenas \emph{et al}. \cite{22,22x} used the KMM model model and ADV  model of GUP to investigate BAU, their results demonstrated that the universe in the GUP framework at non-thermal equilibrium, leading to a non-zero factor $\eta$ for measuring the number of baryonic matter exceeding the antibaryonic matter. Subsequently, Ref.~\cite{23}  have extended this approach to the case of higher-order GUP and showed that the higher order QG correction terms have a significant effect on the baryon asymmetry.

On the other hand, while initial investigations mainly focused on the impact of positive GUP parameters, there has been a growing body of research examining the effects of negative GUP parameters. For instance, the authors in Ref.~\cite{24}  derived the negative GUP parameter (for a GUP on lattice)  for the first time. Then, Scardigli and Casadio derived a negative GUP parameter by assuming that the GUP-corrected Hawking temperature can be obtained through a Wick rotation of an effective Schwarzschild-like metric \cite{24+}. In Ref.~\cite{25}, the properties of sub-Planckian black holes with negative GUP parameters were analysed. Ong pointed out that the Chandrasekhar limit no longer exists in the framework of positive GUP parameter~\cite{26}, whereas the negative GUP parameter solves this contradiction \cite{27}.  In Ref.~\cite{28}, the thermodynamic evolution and phase transition of a static black hole were investigated considering both positive and negative GUP parameters. The findings indicated distinct behaviors in the thermodynamic properties of black holes depending on whether the GUP parameters were positive or negative. Based on the aforementioned research, it is believed that the GUP model with negative parameters is as important for the correction of the physical theory as the case with positive parameters. However, a contradiction arises when the parameter is negative as the GUP no longer includes a minimum length~\cite{29,30}, which contradicts the model-independent existence of the minimum length~\cite{31}. For addressing this issue, Du and Long \cite{32} recently introduced a novel uncertainty principle model, which can be expressed as
\begin{align}
\label{eq1}
\Delta x\Delta p \geq \frac{\hbar }{2}\frac{1}{{1 \pm \left( {{{16{\beta _0}\ell _p^2} \mathord{\left/ {\vphantom {{16{\beta _0}\ell _p^2} {\Delta {x^2}}}} \right.
 \kern-\nulldelimiterspace} {\Delta {x^2}}}} \right)}},
\end{align}
with the deformation parameter ${\beta _0}$ and the Planck length ${\ell_p}$. $\Delta x$ and $\Delta p$ are the uncertainties for position and momentum, respectively. It is important to highlight that in Ref.~\cite{32}, the equation mentioned above is referred to as the GUP. However, it is crucial to note that the RHS of the equation is no longer related to $\Delta x$ but to $\Delta x_0$. Therefore, in accordance with the definition provided in Refs.~~\cite{b5,b5+,b5++}, Eq.~(\ref{eq1})  should be more appropriately labeled as the extended uncertainty principle (EUP). Now, Eq.~(\ref{eq1}) exhibits two primary characteristics: one is the  deformation parameter ${\beta _0}$ (EUP parameter) in the model can be assigned either positive or negative values, with ``$+$" representing the positive EUP parameter and ``$-$" representing the negative EUP parameter. The other one is it maintains a fixed and uniform minimum length ${{\Delta }}{x_{\rm min}} = 4\sqrt {\left| {\rm{\beta_0 }} \right|}\ell _p $, irrespective of whether the parameter is positive or negative. These advantages guarantee the efficacy of the QG effect within the model, enabling us to analyze the impact of both positive and negative EUP parameters on the identical physical system. In addition, if ignoring the higher order correction terms, Eq.~(\ref{eq1}) reduce to $\Delta x\Delta p \geq \hbar {{\left[ {1 + {{16\ell _p^2{\beta _0}} \mathord{\left/ {\vphantom {{16\ell _p^2{\beta _0}} {\Delta {x^2}}}} \right. \kern-\nulldelimiterspace} {\Delta {x^2}}}} \right]} \mathord{\left/  {\vphantom {{\left[ {1 + {{16\ell _p^2{\beta _0}} \mathord{\left/ {\vphantom {{16\ell _p^2{\beta _0}} {\Delta {x^2}}}} \right. \kern-\nulldelimiterspace} {\Delta {x^2}}}} \right]} 2}} \right. \kern-\nulldelimiterspace} 2}$, when comparing with the quadratic GUP (i.e.,  $\Delta x\Delta p \geq {{\hbar \left[ {1 + \beta _0^{{\text{KMM}}}\ell _p^2\Delta {p^2}/{\hbar ^2}} \right]} \mathord{\left/ {\vphantom {{\hbar \left[ {1 + \beta _0^{{\text{KMM}}}\ell _p^2\Delta {p^2}/{\hbar ^2}} \right]} 2}} \right. \kern-\nulldelimiterspace} 2}$), it is found that $\left| {{\beta _0}} \right| \sim \beta _0^{{\text{KMM}}}$.

In this paper, our aim is to integrate the new higher-order EUP with gravitational baryogenesis in order to explore the BAU. By incorporating Eq.~(\ref{eq1}) into the first law of thermodynamics, we derive modified Friedmann equations. These equations reveal that effects of QG play a crucial role in enhancing the energy density and pressure during the radiation-dominated era, thereby disrupting the thermal equilibrium of the universe and ultimately leading to the emergence of a non-zero baryon asymmetry factor $\eta$. Finally, we estimate the bounds of the EUP parameters for both positive and negative cases.

This paper is organized as follows. In section~\ref{sec2}, we briefly review the original scheme of gravitational baryogenesis. In Section\ref{sec3}, by using the new higher-order EUP~(\ref{eq1}) under different parameter signs, we derive modified Bekenstein-Hawking entropy and Friedmann equations. Then, combining the modifications with gravitational baryogenesis, a non-zero time derivative of the Ricci scalar curvature $\dot R$ and a non-zero baryon asymmetry factor $\eta$ are obtained. In section~\ref{sec4}, based on the data of experiments and observations, we constrained the boundaries of both positive and negative deformation parameters. Conclusions and discussion are presented in section~\ref{sec5}. For later convenience, we use the units $\hbar  = c = {k_B} = 1$.

\section{Gravitational baryogenesis in the standard cosmological model}
\label{sec2}
In order to break the CPT dynamically in an expanding Universe, Davoudiasl \emph{et al}. constructed the following interaction \cite{6}:
\label{eq2}
\begin{align}
	\frac{1}{{M_*^2}}\int {{d^4}x\sqrt { - g} \left( {{\partial _\mu }R} \right){J^\mu }} ,
\end{align}
where ${J^\mu }$ denotes the baryon current, ${M_{\rm{*}}}$ is the cut-off scale of the effective theory, $g$ and $R$ stand for the metric and the Ricci curvature scalar, respectively. The baryon asymmetry is usually characterised by the baryon asymmetry
factor (BAF), that is $\eta  = {{{n_B}} \mathord{\left/  {\vphantom {{{n_B}} s}} \right. \kern-\nulldelimiterspace} s}$ with the baryon number density $n_B$ and the entropy density for the universe $s = {{2{\pi ^2}{g_{\text{*}}}{T^3}} \mathord{\left/		{\vphantom {{2{\pi ^2}{g_{\text{*}}}{T^3}} {45}}} \right.\kern-\nulldelimiterspace} {45}}$. Notably, during the expansion of the universe, if the temperature of universe $T$ drops below the critical temperature ${T_D}$ for baryon asymmetric interactions to occur, the BAF can be expressed as \cite{7,11,12,13,14,15,16}
\begin{align}
\label{eq3}
\eta  = \frac{{{n_B}}}{s} \simeq  - {\left.{\frac{{15{g_b}}}{{4{\pi ^2}{g_*}}}\frac{{\dot R}}{{M_*^2T}}} \right|_{{T_D}}},
\end{align}
where $\dot R = {{\partial R} \mathord{\left/ {\vphantom {{\partial R} {\partial t}}} \right. \kern-\nulldelimiterspace} {\partial t}}$ is the time derivative of the Ricci scalar curvature of the universe. ${g_b}$ and ${g_{\rm{*}}}$ are the number of intrinsic degrees of freedom of baryons and the degrees of freedom of particles that contribute to the entropy of universe, respectively.   In the standard cosmological model with the energy density $\rho $ and pressure $p$, the Ricci curvature scalar is given by
\begin{align}
\label{eq4}
R =  - 8\pi G\left( {\rho  - 3p} \right).
\end{align}
Since the matter source in the universe can be seen
as a perfect fluid, the equation of state parameter of standard cosmological model becomes $w = {p \mathord{\left/ {\vphantom {p \rho }} \right. \kern-\nulldelimiterspace} \rho}$, and Eq.~(\ref{eq4}) evolves as
\begin{align}
\label{eq5}
R =  - 8\pi G\rho \left( {1 - 3w} \right).
\end{align}
Notably, when considering the radiation-dominated era, one has $w = {1 \mathord{\left/ {\vphantom {1 3}} \right. \kern-\nulldelimiterspace} 3}$, which leads to the Ricci scalar curvature $R$ and its derivative $\dot R$ vanish, and then resulting in $\eta  = 0$. However, according to many astronomical observations, it is shown that the BAF is not equal to zero, implying that there is more matter than antimatter in the universe. For instance, Particle Data Group gives $\eta  \le 8.6 \times {10^{ - 11}}$ \cite{33}, and BBN shows $3.4 \times {10^{ - 10}} \le \eta  \le 6.9 \times {10^{ - 10}}$ \cite{34}. To solve this problem, we will investigate the gravitational baryogenesis with higher-order EUP corrections.

\section{Gravitational baryogenesis in the framework of higher-order EUP}
\label{sec3}
\subsection{The EUP corrected  entropy }
According to the holographic principle, when a gravitational system absorbs a particle, the area of its apparent horizon and the total energy within it increase. The minimal change of the area ${\rm{\Delta }}A$ can be expressed as \cite{35,36}
\begin{align}
\label{eq6}
\Delta A\sim X m,
\end{align}
where $X$ and $m$ denote the size and mass of the particle, respectively. However, in quantum mechanics, the standard deviation of the $X$ distribution is used to describe the width of a particle's wave packet (the position uncertainty $\Delta x$) and the momentum uncertainty $\Delta p$ is not allowed to be larger than the mass. Thus, Eq.~(\ref{eq6}) can be rewritten as
\begin{align}
\label{eq7}
\Delta A\ge \Delta x  \Delta p	.
\end{align}
Implying that the minimum increment of the area of the gravitational system is limited by the momentum uncertainty $\Delta p$ and the position uncertainty $\Delta x$ of quantum mechanics. According to Eq.~(\ref{eq1}), the momentum uncertainty is obtained as
\begin{align}
\Delta p \geq \frac{\hbar }{2}\frac{1}{{\Delta x \pm {{16\beta_0 \ell_p^2} \mathord{\left/{\vphantom {{16\beta \ell _p^2} {\Delta x}}} \right. \kern-\nulldelimiterspace} {\Delta x}}}}.
\end{align}
For a static spherical gravitational system, the position uncertainty is approximately equal to the radius of the apparent horizon~\cite{37}, which is $\Delta x \approx 2r$. Combining with the above equations, the minimal change of the area is
\begin{align}
\label{9}
\Delta A \ge \chi \tilde \hbar \left( \beta_0  \right),
\end{align}
where $\tilde \hbar \left( \beta_0  \right) = {1 \mathord{\left/  {\vphantom {1 {\left[ {2 + \left( {{{32\pi \beta \ell _p^2} \mathord{\left/  {\vphantom {{32\pi \beta \ell _p^2} A}} \right.  \kern-\nulldelimiterspace} A}} \right)} \right]}}} \right. \kern-\nulldelimiterspace} {\left[ {2 + \left( {{{32\pi \beta_0 \ell_p^2} \mathord{\left/  {\vphantom {{32\pi \beta_0 \ell_p^2 } A}} \right.
\kern-\nulldelimiterspace} A}} \right)} \right]}}$ is the effective Planck constant, and $\chi  = 4\ln 2$ is the calibration factor \cite{38}. When $\beta_0 = 0$, one has $\tilde \hbar \left( \beta_0  \right) = {1 \mathord{\left/		{\vphantom {1 2}} \right.\kern-\nulldelimiterspace} 2}$. Based on the information theory, the minimal increase of entropy is related to the value of the area
\begin{align}
\label{eq10}
\frac{{\rm{d}}S}{{\rm{d}}A} \simeq \frac{{\Delta {S_{\min }}}}{{\Delta {A_{\min }}}} = \frac{1}{{8\tilde \hbar \left( \beta_0  \right)}}.
\end{align}
In the classical limit, the original entropy of a gravitational system can be expressed as ${S_0} = {A \mathord{\left/ {\vphantom {A {4G}}} \right. \kern-\nulldelimiterspace} {4G}}$. Typically, if one wants to modify the entropy, the area $A$ becomes a function of $A$, i.e., $f\left( A \right)$. So when the effect of EUP is considered, the general expression of entropy becomes ${S_0} = {{f\left( A \right)} \mathord{\left/ {\vphantom {{f\left( A \right)} {4G}}} \right. \kern-\nulldelimiterspace} {4G}}$. The relationship between entropy and area is obtained by calculating the derivative of this entropy with respect to area $A$ \cite{39}
\begin{align}
\label{eq11}
\frac{{{\rm d} S}}{{{\rm d} A}} = \frac{{f'\left( A \right)}}{{4G}},
\end{align}
where $f'\left( A \right) = {{{\text{d}}f\left( A \right)} \mathord{\left/ {\vphantom {{{\text{d}}f\left( A \right)} {{\text{d}}A}}} \right. \kern-\nulldelimiterspace} {{\text{d}}A}}$. Now, by comparing Eq.~(\ref{eq10}) with Eq.~(\ref{eq11}), for $\beta_0>0$ case, one has
\begin{align}
\label{eq12}
f'\left( A \right) = \frac{1}{{2\tilde \hbar \left( \beta_0  \right)}} = 1 + \frac{{16 \pi\beta_0 \ell_p^2 }}{A}.
\end{align}
For $\beta_0<0$ case, this relationship becomes
\begin{align}
\label{eq12+}
f'\left( A \right) = \frac{1}{{2\tilde \hbar \left( \beta_0  \right)}} = 1 - \frac{{16 \pi \beta_0 \ell_p^2 }}{A}.
\end{align}
However, when $\beta_0 = 0$, one has $f'\left( A \right) = 1$, which consistent with the standard result in classical limit. Then, by integrating Eq.~(\ref{eq12}), the EUP corrected the entropy with positive parameter is given by
\begin{align}
\label{eq13}
{S_{{\text{EUP}}}} = \int {\frac{{f'\left( A \right)}}{{4G}}{\text{d}}A}  = \frac{A}{{4G}} + \frac{{4\pi }}{G}\beta_0 \ell_p^2 \ln A,
\end{align}
whereas the negative EUP parameter case is
\begin{align}
\label{eq13+}
{S_{{\text{EUP}}}}=\frac{A}{{4G}} - \frac{{4\pi }}{G} \beta_0 \ell_p^2 \ln A.
\end{align}
Clearly, the EUP yields entropy with the parameter $\beta_0$  and a product of logarithmic terms. The incorporation of logarithmic corrections to entropy holds validity in 4 (or 3+1) dimensions. Nevertheless, it is worth noting that in varying dimensions, the modified entropy is recognized to exhibit power law behavior, namely, polynomial form, rather than logarithms \cite{b3,b4}. In the next subsection, we will use Eq.~(\ref{eq13}) and Eq.~(\ref{eq13+}) to derive the modified Friedmann equations.

\subsection{The EUP corrected Friedmann equations }
\label{sec4}
In Refs.~\cite{38,39,40,41,42}, it is found that the Friedman equations can be derived from Bekenstein-Hawking entropy and the first law of thermodynamics. Therefore, based on the modified entropy formula, we will derive the EUP-corrected Friedman equations that deviate from the thermal equilibrium. In the homogeneous and isotropic spacetime, the FRW universe is described by the line element as follows:
\begin{align}
\label{eq14}
{\rm{d}}{s^2} = {h_{\mu \nu }}{\rm{d}}{x^\mu }{\rm{d}}{x^\nu } + {\tilde r^2}\left( {{\rm{d}}{\theta ^2} + {{\sin }^2}\theta {\rm{d}}{\varphi ^2}} \right),
\end{align}
where ${x^\mu } = \left( {t,r} \right),\tilde r = ra\left( t \right)$ with the scale factor $a\left( t \right)$, ${h_{\mu \nu }} = {\rm{diag}}\left[ { - 1,\frac{{{a^2}}}{{\left( {1 - k{r^2}} \right)}}} \right]$ is the two-dimensional metric with the spatial curvature constant $k$ and $\mu ,\nu  = 0,1$ , respectively. According to  ${h^{\mu \nu }}{\partial _u}\tilde r{\partial _\nu }\tilde r = 0$, one can re-express the dynamical apparent of FRW universe as $\tilde r = ar = {\left( {{H^2} + {k \mathord{\left/{\vphantom {k {{a^2}}}} \right.\kern-\nulldelimiterspace} {{a^2}}}} \right)^{ - \frac{1}{2}}}$ with the Hubble parameter $H = {{\dot a} \mathord{\left/ {\vphantom {{\dot a} a}} \right.\kern-\nulldelimiterspace} a}$. Since the matter of the FRW universe can be considered as a perfect fluid, the energy-momentum tensor of the FRW universe is given by
\begin{align}
\label{eq15}
{T_{\mu \nu }} = \left( {\rho  + p} \right){u_\mu }{u_\nu } + p{g_{\mu \nu }},
\end{align}
where ${\rho}$, $p$, ${u_\mu }$ and ${g_{\mu \nu }}$ represent the energy density, pressure, the four velocity of the fluid, and the spacetime metric of FRW universe, respectively. According to the law of conservation of energy-momentum $T_{;\nu }^{\mu \nu } = 0$, the continuity equation can be obtained as
\begin{align}
\label{eq16}
\dot \rho  + 3H\left( {\rho  + p} \right) = 0.
\end{align}
Next, we derive the Friedmann equation from a thermodynamic point of view in terms of the modified entropy~(\ref{eq12+}). According to Ref.~\cite{43}, the first law of thermodynamics for the matter content within the apparent horizon takes the following form
\begin{align}
\label{eq17}
{\rm{d}}E = T{\rm{d}}S + W{\rm{d}}V,
\end{align}
where $E = \rho V$ denotes the total of all the matter energy in the apparent horizon, $V = {{4\pi {{\tilde r}^3}} \mathord{\left/
{\vphantom {{4\pi {{\tilde r}^3}} 3}} \right. \kern-\nulldelimiterspace} 3}$ is the 3-dimensional sphere's volume, and $W = {{\left( {\rho  - p} \right)} \mathord{\left/	 {\vphantom {{\left( {\rho  - p} \right)} 2}} \right.	\kern-\nulldelimiterspace} 2}$ is the density of work. Based on the Eq.~(\ref{eq16}), the differential form of energy can be expressed as
\begin{align}
\label{eq18}
{\rm{d}}E = \rho {\rm{d}}V + V{\rm{d}}\rho  = 4\pi {\tilde r^2}\rho {\rm{d}}\tilde r + \frac{{4\pi {{\tilde r}^3}}}{3}{\rm{d}}\rho,
\end{align}
Then, the temperature of the apparent horizon is $T = {\kappa  \mathord{\left/	{\vphantom {\kappa  {2\pi }}} \right.\kern-\nulldelimiterspace} {2\pi }}$, and the surface gravity for the metric FRW universe is $\kappa  = {{ - \left( {1 - {{\mathop {\tilde r}\limits^.} \mathord{\left/ {\vphantom {{\mathop {\tilde r}\limits^.} {2H\tilde r}}} \right. \kern-\nulldelimiterspace} {2H\tilde r}}} \right)} \mathord{\left/ {\vphantom {{ - \left( {1 - {{\mathop {\tilde r}\limits^.} \mathord{\left/ {\vphantom {{\dot \tilde r} {2H\tilde r}}} \right. \kern-\nulldelimiterspace} {2H\tilde r}}} \right)} {\tilde r}}} \right. \kern-\nulldelimiterspace} {\tilde r}}$ with $\mathop {\tilde r}\limits^. = {{\partial \tilde r} \mathord{\left/ {\vphantom {{\partial \tilde r} {\partial t}}} \right. \kern-\nulldelimiterspace} {\partial t}}$. Thus, the equation for the first law of thermodynamics on the right can be rewritten as
\begin{align}
\label{eq19}
T{\rm{d}}S =  - \frac{1}{{2\pi \tilde r}}\left( {1 - \frac{{\mathop {\tilde r}\limits^. }}{{2H\tilde r}}} \right)\frac{{f'\left( A \right)}}{{4G}}{\rm{dA}},
\end{align}
\begin{align}
\label{eq20}
W{\rm{d}}V = 2\pi {\tilde r^2}\left( {\rho  - p} \right){\rm{d}}\tilde r.
\end{align}
Combining the FRW cosmic dynamics horizon $\tilde r$, and because the radius of the horizon horizon is fixed within an infinitesimal time interval, that is $\mathop {\tilde r}\limits^.  = 0$. Then the first Friedman equation is
\begin{align}
\label{eq21}
- 4\pi G\left( {\rho  + p} \right) = \left( {\dot H - \frac{k}{{{a^2}}}} \right)f'\left( A \right).
\end{align}
Substituting the continuity equation (\ref{eq16}) into the Eq.~(\ref{eq21}), and integrating, the second Friedmann equation is obtained as follows:
\begin{align}
\label{eq22}
\frac{8}{3}\pi G\rho  =  - 4\pi \int {f'\left( A \right)\frac{{{\text{d}}A}}{{{A^2}}}}.
\end{align}
By putting Eq.~(\ref{eq12}) into Eq.~(\ref{eq21}) and Eq.~(\ref{eq22}), the EUP corrected Friedmann equations are given by
\begin{align}
\label{eq23}
- 4\pi G\left( {\rho  + p} \right) = \left( {\dot H - \frac{k}{{{a^2}}}} \right)\left( {1 + \frac{{16\pi \beta_0 \ell_p^2}}{A}} \right),
\end{align}
\begin{align}
\label{eq24}
\frac{8}{3}\pi G\rho  = \frac{{4\pi \left( {A + 8 \pi \beta_0 \ell_p^2 } \right)}}{{{A^2}}} + \mathcal{C}.
\end{align}
In the vacuum energy era, the boundary conditions determine the integration constant $\mathcal{C}$ \cite{22}. When the apparent horizon A tends to infinity, the energy density becomes a cosmological constant $\rho  = {\rm{\Lambda }}$, which leads to  $\mathcal{C} = {{8\pi G\Lambda } \mathord{\left/ {\vphantom {{8\pi G\Lambda } 3}} \right. \kern-\nulldelimiterspace} 3}$. Considering $A = 4\pi {\tilde r^2} = {{4\pi } \mathord{\left/		{\vphantom {{4\pi } {\left( {{H^2} + \frac{k}{{{a^2}}}} \right)}}} \right.
\kern-\nulldelimiterspace} {\left( {{H^2} + \frac{k}{{{a^2}}}} \right)}}$, the modified Friedmann equations becomes
\begin{align}
\label{eq25}
- 4\pi G\left( {\rho  + p} \right) = \left( {\dot H - \frac{k}{{{a^2}}}} \right)\left[ {1 + 4\beta_0 \ell_p^2 \left( {{H^2} + \frac{k}{{{a^2}}}} \right)} \right],
\end{align}
\begin{align}
\label{eq26}
\frac{8}{3}\pi G\left( {\rho  - \Lambda } \right) = \frac{{\frac{{4\pi }}{{\left( {{H^2} + {k \mathord{\left/{\vphantom {k {{a^2}}}} \right.
\kern-\nulldelimiterspace} {{a^2}}}} \right)}} + 8\pi\beta_0 \ell_p^2 }}{{{{\left[ {{{4\pi } \mathord{\left/{\vphantom {{4\pi } {\left( {{H^2} + \frac{k}{{{a^2}}}} \right)}}} \right.		 \kern-\nulldelimiterspace} {\left( {{H^2} + \frac{k}{{{a^2}}}} \right)}}} \right]}^2}}}.
\end{align}
In line with our research objective, our focus will be on studying a flat universe that is dominated by radiation. Therefore, the cosmological constant $\Lambda$ and the spatial curvature constant $k$ should reduce to zero\cite{22}, resulting in the modified equations taking the following form
\begin{align}
\label{eq27}
- 4\pi G\left( {\rho  + p} \right) = \dot H\left( {1 + 4\beta_0 \ell_p^2{H^2}} \right),
\end{align}
\begin{align}
\label{eq28}
\frac{8}{3}\pi G\rho  = {H^2}\left( {1 + 2\beta_0 \ell_p^2{H^2}} \right).
\end{align}
It needs to be noted that Eq.~(\ref{eq27}) and Eq.~(\ref{eq28}) are valid on the basis of $\beta_0 > 0$. With the same derivation, one can obtain the results for the case $\beta_0  < 0$ as follow
\begin{align}
\label{eq27+}
- 4\pi G\left( {\rho  + p} \right) = \dot H\left( {1 - 4 \beta_0 \ell_p^2{H^2}} \right),
\end{align}
\begin{align}
\label{eq28+}
\frac{8}{3}\pi G\rho  = {H^2}\left( {1 - 2 \beta_0 \ell_p^2{H^2}} \right).
\end{align}
The aforementioned findings demonstrate that the EUP has a distinct impact on the energy density and pressure in the early universe, irrespective of whether the EUP parameters are positive or negative. However, when $\beta_0 = 0$, the modifications return to the original cases. In the subsequent section, they will be used to break thermodynamic equilibrium and explain baryon asymmetry in the early universe.

\subsection{The EUP corrected baryon asymmetry factor}
From the above, it is clear that in the standard cosmological model, the Ricci curvature scalar can be expressed by the energy density ${{\rho }}$ and the pressure ${p}$. In order to investigate the universe's deviation of thermal equilibrium, it is indispensable to  re-express the  ${{\rho }}$ and   ${{p}}$ as $\rho _{{\rm{EUP}}} = {\rho _0} + {{\Delta }}\rho $ and $ p_{{\text{EUP}}} = {p_0} + {{\Delta }}p$, where $\rho _0$ and ${p_0}$ denote the original energy density and pressure at thermal equilibrium, respectively. ${\rm{\Delta \rho }}$ and  ${\rm{\Delta }}p$ are  correction quantities.    Substituting $\rho _{{\rm{EUP}}}$ and $p_{{\text{EUP}}}$ into the Eq.~(\ref{eq27}) and Eq.~(\ref{eq28}),  considering $\dot H =  - 4\pi G\left( {{\rho _0} + {p_0}} \right)$ and ${H^2} = {{8\pi G{\rho _0}} \mathord{\left/
{\vphantom {{8\pi G{\rho _0}} 3}} \right. \kern-\nulldelimiterspace} 3}$, one yields
\begin{align}
\label{eq29}
{\rho _{{\rm{EUP}}}} = {\rho _0} + \frac{{16}}{3}G\pi \beta_0 \ell_p^2 {\rho _0}^2,
\end{align}
\begin{align}
\label{eq30}
{p_{{\text{EUP}}}} = w{\rho _0} + \frac{{16G\pi \beta_0 \ell_p^2 }}{3}\left( {1 + 2w} \right){\rho _0}^2.
\end{align}
The results demonstrate that the QG effect in the higher-order EUP can cause changes in energy density and pressure in the radiation-dominated era, breaking the universe's thermal equilibrium and fulfilling the third Sakharov condition. Next, we substitute the corrected energy density and pressure into Eq.~(\ref{eq5}) to obtain the corrected Ricci curvature scalar:
\begin{align}
\label{eq31}
{R_{\rm EUP}}=&\frac{8}{3}G\pi {\rho _0}\left[ {32G\pi \beta_0 \ell_p^2 {\rho _0} - 3 +w\left( {9 + 96G \pi \beta_0 \ell_p^2 {\rho _0}} \right)} \right].
\end{align}
Both energy density and pressure are functions with respect to time, and using the continuity equation, the time derivative of the modified Ricci curvature scalar is obtained as
\begin{align}
\label{eq32}
{\dot R_{{\text{EUP}}}} & =  - 16\sqrt {\frac{2}{3}} {\pi ^{\frac{3}{2}}}\left( {1 + w} \right){\left( {G{\rho _0}} \right)^{\frac{3}{2}}}
\nonumber \\
&\times \left[ {64G \pi \beta_0 \ell_p^2 {\rho _0} - 3 + 3w\left( {3 + 64G \pi \beta_0 \ell_p^2 {\rho _0}} \right)} \right].
\end{align}
Here we set $w = 1/3 $ because the universe  is in a radiation-dominated era, so the above equation becomes:
\begin{align}
\label{eq33}
{\dot R_{\rm EUP}} =  - \frac{{8192}}{3}\sqrt {\frac{2}{3}} G {\pi ^{\frac{5}{2}}}\beta_0 \ell_p^2 {\rho _0}{(G{\rho _0})^{\frac{3}{2}}}.
\end{align}
This is the EUP corrected the time derivative of the Ricci curvature scalar in the radiation-dominated era, substituting Eq.~(\ref{eq33}) into the expression of Eq.~ (\ref{eq3}), the modified BAF is given by
\begin{align}
\label{eq34}
{\eta _{\rm EUP}} = \frac{{10240 \sqrt {\frac{{2\pi }}{3}} \beta_0 \ell_p^2 {{(G{\rho _0})}^{\frac{5}{2}}}{g_b}}}{{{g_{\rm{*}}}M_{\rm{*}}^2{T_D}}}.
\end{align}
Similarly, by using the Eq.~(\ref{eq27+}) and Eq.~(\ref{eq28+}), the expression of BAF for $\beta_0<0$ becomes
\begin{align}
\label{eq35}
{\eta _{{\text{EUP}}}} =  - \frac{{10240\sqrt {\frac{{2\pi }}{3}} \beta_0 \ell_p^2 {{\left( {G{\rho _0}} \right)}^{\frac{5}{2}}}{g_b}}}{{{g_*}M_*^2{T_D}}}.
\end{align}
Obviously, Eq.~(\ref{eq34}) and Eq.~(\ref{eq35}) are on longer zero, which are associated with the deformation parameter ${\rm{\beta_0 }}$. Our results demonstrate that the EUP with positive and negative parameters can generate baryon asymmetry in the radiation-dominated era.

\section{The bounds of the deformation parameters $\beta_0$}
\label{sec4}
Besides the numerous theoretical studies on the EUP, a popular study direction in QG phenomenology focuses on attempting to quantify the size of EUP corrections by constraining the deformation parameter~\cite{44,45,46,47,47+,47a+,47b+}. The deformation parameters are theoretically constantly assumed to be 1, allowing the QG effect to be valid when the energy is close to the Planck scale. The bounds of the deformation parameters can be determined with experimental data and observations when the assumption is ignored. Therefore, in this section, we will compare the theoretical results with the observed results to obtain the bounds of deformation parameter. Then, we need to replace the gravitational mass with the Planck mass $G = {1 \mathord{\left/ {\vphantom {1 {M_p^2}}} \right. \kern-\nulldelimiterspace} {M_p^2}}$, with $M_p \sim 1.22 \times {10^{28}}{\rm eV}$. The density at thermal equilibrium can be expressed as ${\rho _0} = {{\pi {g_{\text{*}}}{T_D}^4} \mathord{\left/	{\vphantom {{\pi {g_{\text{*}}}{T_D}^4} {30}}} \right.	\kern-\nulldelimiterspace} {30}}$, when the deformation parameter takes positive, the BAF is rewritten as
\begin{align}
\label{eq40}
{\eta _{{\text{EUP}}}} = \frac{{512{\pi ^3}\beta_0 \ell_p^2 {g_b}}}{{135\sqrt 5 {g_{\text{*}}}M_{\text{*}}^2{T_D}}}{\left( {\frac{{{g_{\text{*}}}{T_D}^4}}{{{M_p}^2}}} \right)^{\frac{5}{2}}}.
\end{align}
To further obtain the bounds of the EUP parameter, we set ${M_*} = {{{M_p}} \mathord{\left/ {\vphantom {{{M_{{p}}}} {\sqrt {8\pi } }}} \right.
\kern-\nulldelimiterspace} {\sqrt {8\pi }}}$, $T_D \sim 2\times {10^{25}}{\rm{eV}}$, ${{g_*}} = 106$, ${{g_b}} = 1$. Then, solving Eq.~(\ref{eq40}), the deformation parameter $\beta_0 > 0$ can be expressed as
\begin{align}
\label{eq41}
{{\rm{\beta }}_0} = 8.1 \times {10^{18}}{\eta _{\rm EUP}}.
\end{align}
When the deformation parameter takes negative, the corresponding constraint is obtained in the same way:
\begin{align}
\label{eq42}
-{{\rm{\beta }}_0} = 8.1 \times {10^{18}}{\eta _{\rm EUP}}.
\end{align}
Obviously, the above equations show that the bounds of the deformation parameter ${{\rm{\beta }}_0}$  are determined by the observation ${\eta _{\rm EUP}}$. We list the range of ${\eta}$ given by astronomical observations over the last four decades, which can be used to constrain the bounds of higher-order generalized uncertainty principle parameters, as follows in Table~\ref{tab1}.
\begin{table*}[htbp]
\centering
\renewcommand\arraystretch{2.0}
\caption{\label{tab1} The range of $\eta$ and the bounds of ${{\rm{\beta}}_0}$ from data source.}
\resizebox{1.0\linewidth}{!}{
\begin{tabular}{cccc}
\hline
Data source & $\eta $ & ${{\rm{\beta }}_0} > 0$ & ${{\rm{\beta }}_0} < 0$\\
\hline
Particle Data Group\cite{33} & $\eta  \le 8.6 \times {10^{ - 11}}$ &	${{\rm{\beta }}_0} \le 6.9 \times {10^{8}}$ &	${{\rm{\beta }}_0} \ge  - 6.9 \times {10^{8}}$ \\
BBN\cite{34} & $3.4 \times {10^{ - 10}} \le \eta  \le 6.9 \times {10^{ - 10}}$ & $2.7 \times {10^{9}} \le {{\rm{\beta }}_0} \le 5.5 \times {10^{9}}$ &$ - 5.8 \times {10^{9}} \le {{\rm{\beta }}_0} \le  - 2.9 \times {10^{9}}$ \\
Planck observations\cite{48} & 	$\eta  \le 6.2 \times {10^{ - 10}}$ & 	${{\rm{\beta }}_0} \le 5.0 \times {10^{9}}$ & ${{\rm{\beta }}_0} \ge  - 5.0 \times {10^{9}}$ \\
Deuterium and 3He abundances\cite{49} & $5.7 \times {10^{ - 11}} \le \eta  \le 9.9 \times {10^{ - 11}}$ & 	$4.6 \times {10^{8}} \le {{\rm{\beta }}_0} \le 8.0 \times {10^{8}}$ & 	$ - 8.0 \times {10^{8}} \le {{\rm{\beta }}_0} \le  - 4.6 \times {10^{8}}$ \\
Acoustic peaks in CMB measured by WMAP\cite{50} & $\eta  \le 6.3 \times {10^{ - 10}}$ & ${{\rm{\beta }}_0} \le 5.1 \times {10^{9}}$ & 	${{\rm{\beta }}_0} \ge  -5.1\times {10^{9}}$ \\
Deuterium and Hydrogenium abundance \cite{51,52} & $5.7 \times {10^{ - 10}} \le \eta  \le 6.7 \times {10^{ - 10}}$ & 	$4.6 \times {10^{9}} \le {{\rm{\beta }}_0} \le 5.4 \times {10^{9}}$ & 	$ - 5.4 \times {10^{9}} \le {{\rm{\beta }}_0} \le  - 4.6 \times {10^{9}}$ \\
\hline
\end{tabular}}
\end{table*}

In Table~\ref{tab1}, by utilizing the data from different experiments and observation results, it is found that the bounds of positive deformation parameter is  $4.6 \times {10^{8}}\sim 5.5 \times {10^{9}}$, while the bounds of negative deformation parameter is $ -5.5 \times {10^{9}} \sim - 4.6 \times {10^{8}}$. Furthermore, we would like to compare our results with those in Ref.~\cite{22}. By subjecting both models to the same conditions, specifically the Deuterium and 3He abundances~\cite{49}, one can directly compare their respective outcomes. The result obtained from the KMM model is expressed as  $\beta _0^{{\text{KMM}}} =  - 4.63 \times {10^{18}}\eta $, which subsequently yields a constraint on the deformation parameter, i.e.,   $- 4.58 \times {10^8} \leq \beta _0^{{\text{KMM}}} \leq - 2.64 \times {10^8}$. It is easy found that the absolute values of our results $\left| {{\beta _0}} \right|$  have the same order of magnitude as  $\beta _0^{{\text{KMM}}}$, which is consistent with the relationship between the two deformation parameter that we presented in section 1, that is  $\left| {{\beta _0}} \right| \sim \beta _0^{{\text{KMM}}}$. However, it should be noted that the KMM model can only give bound for negative parameter, while ours can give both positive and negative parameter cases.

\section{Conclusion}
\label{sec5}
In the present work, we have investigated the gravitational baryogenesis mechanism within the framework of a new higher-order  EUP with a fixed and uniform minimum length in the presence of positive and negative deformation parameters. Our analysis involved deriving the corrected Bekenstein-Hawking entropy and Friedmann equations by combining the higher-order EUP with the first law of thermodynamics. These modifications allowed us to explore the thermodynamic equilibrium of the universe during the radiation era, considering the effects of QG. Finally, we calculated the EUP-corrected baryon asymmetry factor, denoted as $\eta$. Our results revealed that the EUP plays an important role in modifying the pressure ${p}$ and energy density $\rho$ terms in the Friedmann equations, thus disrupting the thermal equilibrium of the universe. This modification addresses a limitation in the original theory by satisfying Sakharov's third condition. Moreover, we observed that the EUP introduces non-zero values for the Ricci scalar $R$ and its time derivative $\dot{R}$, leading to a non-zero $\eta$. This indicates that the EUP can effectively generate baryon asymmetry during the radiation-dominated era.

Furthermore, by comparing our theoretical results with observational data, we have successfully constrained the positive EUP parameter to the range of  $4.6 \times {10^{8}}\sim 5.5 \times {10^{9}}$, and the negative deformation parameter to the range of $ -5.5 \times {10^{9}} \sim - 4.6 \times {10^{8}}$. It is well known that current research has focused on EUP with positive parameter, whereas EUP with negative deformation parameters has received relatively less attention. It demonstrates that EUP with negative deformation parameters can exert the same influence on baryon asymmetry as EUP with positive deformation parameters. Moreover, we have successfully demonstrated the feasibility of constraining both positive and negative deformation parameters using observational data. In future research, our objective is to integrate the newly proposed EUP~(\ref{eq1}) into various other physical theories. We plan to explore its implications in fields such as the Chandrasekhar limit, black hole thermodynamics, and the associated phase transitions. By incorporating the EUP model into these diverse contexts, we aim to gain a deeper understanding of its effects and potential applications. Additionally, we will endeavor to refine the constraints on the parameters of the EUP model by conducting rigorous analysis of experimental data. This data-driven approach will allow us to further validate the model's predictions and assess its accuracy and reliability.

\end{document}